\documentclass[prd,twocolumn,aps,psfig,showpacs,nofootinbib,nobibnotes,superscriptaddress,preprintnumbers]{revtex4}
\usepackage{graphicx}
\usepackage[figuresright]{rotating}
\usepackage{amssymb}
\usepackage{bm}
\usepackage{epsfig}
\usepackage{color}

\newcommand{\beq}{\begin{eqnarray}}
\newcommand{\eeq}{\end{eqnarray}}

\newcommand{\DD}{{\rm{D}}}

\newcommand{\Fig}[1]{Figure~\ref{#1}}

\begin{document}

\title{Evolution of Primordial Magnetic Fields from Phase Transitions}

\date{\today, $ $Revision: 1.47 $ \! $}
\preprint{NORDITA-2012-96}

\author{Tina Kahniashvili}
\email{tinatin@phys.ksu.edu} \affiliation{McWilliams Center for
Cosmology and Department of Physics, Carnegie Mellon University,
5000 Forbes Ave, Pittsburgh, PA 15213, USA}
\affiliation{Department of Physics, Laurentian University, Ramsey
Lake Road, Sudbury, ON P3E 2C,Canada} \affiliation{Abastumani
Astrophysical Observatory, Ilia State University, 3-5 Cholokashvili
Ave., Tbilisi, 0160, Georgia}

\author{Alexander G. Tevzadze}
\email{aleko@tevza.org} \affiliation{Faculty of Exact and Natural
Sciences, Tbilisi State University, 1 Chavchavadze Ave., Tbilisi,
0128, Georgia}

\author{Axel Brandenburg}
\email{brandenb@nordita.org} \affiliation{Nordita,
KTH Royal Institute of Technology and Stockholm University,
Roslagstullsbacken 23, 10691 Stockholm, Sweden}
\affiliation{Department of Astronomy, Stockholm University, 10691
Stockholm, Sweden}

\author{Andrii Neronov}
\email{Andrii.Neronov@unige.ch} \affiliation{ISDC Data Centre for
Astrophysics, Ch. d'Ecogia 16, 1290 Versoix,
Switzerland}\affiliation{Geneva Observatory, Ch. des Maillettes
51, 1290 Sauverny, Switzerland}


\begin{abstract}
We consider the evolution of primordial magnetic fields generated
during cosmological, electroweak or QCD, phase transitions.
We assume that the magnetic field generation can be described as an
injection of magnetic energy to cosmological plasma at a given scale
determined by the moment of magnetic field generation.
A high Reynolds number ensures strong coupling between magnetic field
and fluid motions.
The subsequent evolution of the magnetic field is governed by decaying
hydromagnetic turbulence.
Both our numerical simulations and a phenomenological description allow
us to recover ``universal'' laws for the decay of magnetic energy and the
growth of magnetic correlation length in the turbulent (low viscosity) regime.
In particular, we show that
during the radiation dominated epoch, energy and correlation
length of non-helical magnetic fields scale as conformal time to the
powers $-1/2$ and $+1/2$, respectively. For helical magnetic
fields, energy and correlation length scale as conformal time to
the powers $-1/3$ and $+2/3$, respectively. The universal decay law
of the magnetic field implies that the strength of magnetic field
generated during the QCD phase transition could reach  $\sim 10^{-9}$\,G
with the present day correlation length $\sim 50$~kpc. The fields
generated at the electroweak phase transition could be as strong as
$\sim 10^{-10}$~G with correlation lengths reaching $\sim 0.3$\,kpc.
These values of the magnetic fields are consistent with the
lower bounds of the extragalactic magnetic fields.
\end{abstract}



\maketitle

\section{Introduction}
Astronomical observations show that galaxies have magnetic fields
with a component that is coherent over a large fraction of the
galaxy with field strengths of the order of $10^{-6}$~G; see
Refs.~\cite{Widrow,Vallee,beck} and references therein.  Understanding the
origin of these fields is one of the challenging questions of modern
astrophysics.  Generally speaking, there are two popular scenarios.
The first one envisages the generation of magnetic fields through
different astrophysical mechanisms. More precisely, it is assumed
that an initially tiny magnetic field is produced through
a battery mechanism \cite{Kulsrud:2007an}.
The correlation length of such a field is limited by galactic length scales.
The second scenario to explain the
origin of the magnetic field in galaxies and clusters presumes that
the observed magnetic fields were amplified from cosmological
weak seed magnetic fields \cite{Kandus:2010nw}. In this case the
correlation length of such a seed field might be as large as the horizon
scale today if we admit that the field has been generated during
inflation. There are different possibilities for seed magnetic field
amplification ranging from a magnetohydrodynamic (MHD) dynamo to the adiabatic
compression of magnetic field lines during structure
formation \cite{Beck:1995zs,LSS2011}.

Galactic magnetic fields are usually measured through the induced
Faraday rotation effect \cite{Vallee,beck} and, as mentioned above,
the field magnitude is of the order of a few $10^{-6}\,$G at typical scales
of $10\,$kpc. The primordial magnetic energy density contributes to the
radiation field, and thus the big bang nucleosynthesis (BBN) bound
implies $\Omega_B h_0^2 \leq 2.4\times 10^{-6}$ \cite{grasso} if
the magnetic field has been generated prior to BBN. If the correlation
length of the magnetic field is much larger than $\lambda_B\gtrsim 1$~Mpc,
smaller limits on the magnetic field energy density arise from the
cosmological data, making $B_{\rm max} \leq $ a few $10^{-9}$ G; see
\cite{ktspr10} and references therein.
A correlation length-dependent lower limit on magnetic fields in the
intergalactic medium  (IGM) could be derived from gamma-ray
observations of blazars \cite{neronov,limit2,dolag2,dermer}. In the limit
of large correlation lengths, $\lambda_B\gtrsim 1$\,Mpc, the bound is
at the level of $10^{-17}$\,G (see also Ref.~\cite{Arlen:2012iy} for a
discussions on possible uncertainties in the measurements of blazar spectra).

It is possible, in principle, that
the IGM magnetic fields are  of primordial origin. Another
possibility is that the fields are spread through the IGM by
outflows from galaxies at late stages of the evolution of the
Universe. To distinguish between these two possibilities, it is
important to identify measurable characteristics of the IGM magnetic
fields which are different in the two cases and to study the possibility
of measuring such characteristics.

In this paper we consider the observational properties of IGM
magnetic fields expected if the fields originate from
cosmological phase transitions (PT) such as the electroweak (EW) and
QCD PTs \cite{beo96,phase}.
Some scenarios of magnetic field generation during EWPT or QCDPT also
produce magnetic helicity \cite{phase-hel}.
We follow the evolution of fields from
the epoch right after the magnetogenesis up to the present day
epoch. Our approach is different from that adopted in the previous
studies, which mostly  concentrated on the analysis of the range of
possible field strengths at a pre-defined scale of interest (e.g.\
1\,Mpc). Instead, we study the evolution of field
characteristics that are most relevant for measurements using
radio and gamma-ray astronomy. Specifically, we are
interested in the evolution of the magnetic energy density $\rho_B$,
which determines the characteristic field strength $B^{\rm
(eff)}=\sqrt{8\pi \rho_B}$, and the characteristic correlation
scale (integral scale) $\lambda_M$ at which the field reaches
the strength $B^{\rm (eff)}$.

The present day integral scale depends on the temperature $T_\star$
and the number of relativistic degrees of freedom $g_\star$ at the
moment when the primordial magnetic field  is generated. These
parameters determine the maximal allowed value of the magnetic
energy density injected in the PT plasma, as well as the initial
correlation length of the magnetic field \cite{ktr11}. We do not
separately consider the effect of helicity transfer related to the
chiral anomaly, which might be important in the presence of strong
magnetic fields at temperatures above 10--100\,MeV
\cite{boyarsky}. Such a transfer could be considered as part of
the magnetogenesis process which could persist all the way down to
the temperature scale of the QCD phase transition. We only use
fundamental physical laws, such as conservation of energy, and how
the magnetic field interacts with the cosmological plasma through
MHD turbulence, and do not make any assumption about the physical
processes responsible for the primordial magnetic field generation.

In Sec.\ II we give an overview of the spatial and temporal
characteristics of the primordial magnetic field. The results of our
analysis are presented in Sec.\ III, where we discuss the evolution
of the magnetic field. Conclusions are presented in Sec.\ IV. We
employ natural units with $\hbar = 1 = c$ and gaussian units for
electromagnetic quantities.

\section{Model Description}

\subsection{Effective Magnetic Field Characteristics}

We assume that the phase transition-generated magnetic fields
satisfy the causality condition \cite{hogan,beo96,cd01}. The maximal
correlation length $\xi_{\rm max}$ for a causally generated primordial magnetic field cannot
exceed the Hubble radius at the time of generation, $H_\star^{-1}$.
Hence $\gamma= \xi_{\rm max}/ H_\star^{-1} \leq 1$, where $\gamma$ can
be associated with the number of primordial magnetic field  bubbles
within the Hubble radius, $N \propto \gamma^3$. The comoving length
(measured today) corresponding to the Hubble radius at generation is
inversely proportional to the corresponding PT temperature
$T_\star$,
\begin{equation}
\lambda_{H_\star} = 5.8 \times 10^{-10}~{\rm
Mpc}\left(\frac{100\,{\rm GeV}}{T_\star}\right)
\left(\frac{100}{g_\star}\right)^{{1}/{6}}, \label{lambda-max}
\end{equation}
and is equal to 0.5 pc for the QCDPT (with $g_\star=15$ and
$T_\star=0.15$ GeV) and $6 \times 10^{-4}$ pc for the electro weak phase transitions (EWPT) (with
$g_\star=100$ and $T_\star =100$ GeV), and the comoving primordial magnetic field
correlation length $\xi_{\rm max} \leq \lambda_H$. This inequality
assumes {\it only} the expansion of the Universe without accounting
for MHD turbulence (free turbulence decay or an inverse cascade
if a helical primordial magnetic field is present).
We also note that the number of PT bubbles within the Hubble radius
is around 6 ($\gamma \simeq 0.15$) for QCDPT and around 100
($\gamma \simeq 0.01$) for EWPT. So the maximal correlation length
$\xi_{\rm max}$ is equal to 0.08\,pc for QCDPT and $6 \times 10^{-6}$\,pc
for EWPT.

The maximal value of the primordial magnetic field energy density
must satisfy the BBN bound, i.e.\ the total energy density of the
primordial magnetic field  at nucleosynthesis $\rho_B$ (where $a$
should not exceed $10\%$ of the radiation energy density $\rho_{\rm
rad}(a_{\rm N})$. In any case, the magnetic field is generated by
mechanical motion of charged particles, so that its energy density
could constitute only a fraction of matter energy density, which is
at most comparable to the radiation energy density in the radiation
dominated Universe. Note that the maximal value of the effective
magnetic field  is independent of the temperature at generation
$T_*$, and depends only very weakly on the number of relativistic
degrees of freedom at the moment of generation.

In what follows we are mostly interested in the evolution of the
energy density of the magnetic field and the length scale which gives
the dominant contribution to the energy density (the ``integral
scale'') during the course of cosmological evolution.  Taking this into
account we adopt the following idealizing approximation.
We generate an initial primordial magnetic field by solving the MHD equations
for a certain time during which an external electromagnetic force is applied
that is proportional to a delta function that peaks at the characteristic
scale $k_0=2\pi/\xi_{0}^{-1}$. This corresponds to a magnetic field with
correlation length $\xi_0$. In this approximation the
characteristic magnetic field strength at the scale $\xi_0$ is
$B^{\rm (eff)}=\sqrt{8\pi \rho_B}$. We justify our assumption that
$\xi_0$ should be identified with the size of the largest magnetic
eddies by noting that the primordial magnetic field is involved in MHD
processes driven by turbulence. It is natural to assume that the
typical length scale of the magnetic field generated during the PTs
is determined by the PT bubble size.

The dynamical evolution of the coupled magnetic field--matter system
leads to a spread of magnetic field over a range of scales. The
resulting power spectrum of the magnetic field at small wavenumbers
(or, equivalently, large distance scales) has the form of a
power-law $P_M(k)= {E_M\!(k)}/{(4\pi k^2) } = A k^{n_B}$ with a
normalization constant $A$ and a slope\footnote{In general the
magnetic field spectrum is determined through the Fourier transform
$F_{ij}^M({\bf k})$ of the two-point correlation function of the
magnetic field, $\langle B_i({\bf x}) B_j({\bf x}+{\bf r}) \rangle$,
with the spectral function \cite{my75}
\begin{eqnarray}
F_{ij}^M\!({\bf k}) =  \label{eq:4.1}
P_{ij}({\bf k}) \frac{E^M\!(k)}{4\pi k^2}  + i
\varepsilon_{ijl} {k_l} \frac{H^M\!(k)}{8\pi k^2}.
\end{eqnarray}
Here  $P_{ij}({\bf k}) = \delta_{ij}-{k_i
k_j}/{k^2}$, $\varepsilon_{ijl} $ is the antisymmetric tensor,
$E^M\!(k)$ and $H^M\!(k)$ are the magnetic energy and helicity spectra.} $n_B$.
In particular, a white noise power spectrum corresponds to
$n_B=0$ \cite{hogan}, while the Batchelor spectrum corresponds to
$n_B=2$ \cite{cd01}.
The power law spectrum extends up to a time-dependent integral
scale $\xi_M$ above which the power contained in the magnetic
field decreases rapidly due to turbulent decay and/or viscosity
damping.

Several previous studies, see \cite{sub} and
references therein, describe the primordial magnetic field in terms
of a smoothed (over length scale $\xi$) magnetic field
$B_\xi$ with $B_\xi^2 = \langle B_i({\bf x}) B_i({\bf x})\rangle |_{\xi}$.
Knowing $B^{(\rm eff)}$ and the slope of the
power spectrum, one can calculate the strength of the smoothed
magnetic field at any scale of interest $\xi$ \cite{ktr11}:
\begin{equation}
B_\xi=\frac{B^{(\rm eff )}\sqrt{\Gamma(n_B/2+5/2)}}{(
\xi/\xi_M)^{(n_B+3)/2}}. \label{b-eff}
\end{equation}
{The smoothed magnetic field might be of interest in the context of
certain problems. For example, the strength of a magnetic field
smoothed over a scale $\xi\sim 1$\,Mpc is considered to be relevant
in the context of seed magnetic fields for galactic dynamos. It is,
however, important to note that} $B_\xi$ is strongly $n_B$
dependent for a given value of $\xi$.  In particular, for
causally-generated magnetic fields with $n_B \geq 0$ there is
significant ``magnetic power'' only at small scales, and for
$\xi \simeq 1$ Mpc the value of $B_{1{\rm Mpc}}$ is extremely
small \cite{caprini01}. {At the same time, it does not imply that
the magnetic field itself is weak. In fact, } it could be as strong
as $B^{(\rm eff)} \simeq 10^{-6}-10^{-7}$~G, close to the bound
imposed by the BBN  \cite{ktr11}.  Only in the case of a scale
invariant magnetic field with $n_B \rightarrow -3$, generated for
example during inflation \cite{ratra}, $B_\xi $ is independent
of $\xi$ and $n_B$, and is equal to $B^{\rm (eff)}$.
Note that sometimes (see, e.g., Eq.~(8) in \cite{cdk04}], instead
of calculating $B_\xi$ through $B^{({\rm eff})}$, the smoothed value
of the magnetic field is determined through the normalization
constant $A$ of the power spectrum as
$B_\xi^2 = A \Gamma (n_B/2 +3/2)/(\xi_M)^{(n_B+3)}/(2\pi)^2$.

\subsection{Phenomenological description of the magnetic field decay in the free turbulence regime}

After generation, the evolution of the primordial magnetic field is
a complex process affected by MHD as well as by the expansion of the
Universe
\cite{a1,s1,b1,b2,campanelli2,jedamzik,campanelli,kbtr10,jedamzik2}.
In our description, to account for the expansion of the Universe we
make use of the fact that conformal invariance allows for a
description of MHD processes in the early Universe by simply
rescaling all physical quantities in terms of their comoving values
and using the conformal time $\eta$ \cite{beo96}. After this
procedure the MHD equations include the effects of the expansion
while retaining their conventional flat spacetime form.

The magnetic evolution process strongly depends on initial
conditions, as well as on the physical conditions of the primordial
plasma. We need to determine the scaling laws for the following
magnetic field characteristics: (i) magnetic energy density, (ii)
correlation length, and (iii) magnetic helicity. The magnetic energy
and magnetic helicity spectra are related through the realizability condition,
$|H_M\!(k,\eta)| \leq 2 E_M\!(k,\eta)/k$ \cite{b1}. For the total
magnetic energy ${\mathcal E}_M(t)=\int E_M\!(k,\eta)\, dk$ and
helicity ${\mathcal H}_M(\eta)=\int H_M\!(k,\eta)\, dk$ we get
\begin{equation}
{\mathcal H}_M(\eta) \leq 2 \xi_M(\eta) {\mathcal E}_M(\eta),
\end{equation}
where
\begin{equation}
\xi_M(\eta) \equiv {\mathcal E}^{-1}_M(\eta) \int E_M(k,\eta)k^{-1} dk
\end{equation}
is the {comoving} magnetic eddy correlation length (which
corresponds to the physical integral scale $\lambda_M$), initially
set by the temperature at the magnetic field generation moment
$\xi_{M,{\rm in}} = \lambda_0 =\gamma \lambda_{H_\star}$; see
Eq.~(\ref{lambda-max}), and is independent of the presence of
magnetic helicity.

Both helical and non-helical magnetic fields experience large-scale
MHD decay resulting in an increase of the correlation length with a
corresponding decrease in the magnetic energy density at large
scales; for a review see \cite{dav,b1}. The time rate of this
process depends strongly on the presence of magnetic helicity
\cite{a1,s1}. Taking this into account, we consider the cases of
helical and non-helical magnetic fields separately.

As noted above, the initial magnetic field configuration is
given by a sharply peaked spectral energy density. The coupling
between primordial magnetic field and plasma, which ensures the
spreading of the fixed scale primordial magnetic field over a wide
range of length scales, forms a modified magnetic field spectrum
within a few turnover times (see Ref.~\cite{tkbk12} for more
details). The final realization of the spectrum is given by the
Batchelor spectrum, $n_B=2$. Our numerical simulation results are in
perfect agreement with the ``causality constraint'' that in the cosmological
context has been discussed in Ref.~\cite{beo96}, and studied through analytical
approach in Ref.~\cite{cd01}\footnote{A recent study based on semi-analytical
calculations \cite{jedamzik2012} showed the same shape for phase
transition-generated magnetic fields. In laboratory plasma as
well as in numerical simulations, the resulting magnetic
field spectrum at large scales can be given by a white noise spectrum,
$n_B=0$ \cite{hogan} or even by a flatter Kazantsev spectrum, $n_B=-1/2$
\cite{b1}. However, here we consider the case of a cosmological
magnetic field for which the correlation length is strongly limited by the
Hubble horizon.} so that the power contained in the large-scale modes
is very small, while the total magnetic energy
density is sufficiently large.
MHD processes also are
responsible for the generation of fluid perturbations when an initial
magnetic field is present. These processes finally result in
equipartition between magnetic and kinetic energy densities
\cite{kbtr10,kbcrt12}. In contrast to the magnetic field, the
velocity field has a white noise spectrum, i.e.\
$P_K=E_K/(4\pi k^2)=A_Kk^{n_K}$ with $n_K=0$ due to the possible
presence of longitudinal modes. To describe adequately the
evolution of fluid motions coupled to the magnetic field we need to
solve the complete set of MHD equations; see below.

Below we present a phenomenological description of the MHD
decay laws at large scales.
In the case of helical fields we mostly follow the description
presented in Secs.~4.2.3 and 5.3.2 of Ref.~\cite{b1} and Sec.~7.3.4 of
\cite{bII}. On scales below $\xi_M$, magnetic power is
transferred to smaller scales via the so-called {\it direct cascade} by
turbulence until it is finally damped at the smallest scale
$\xi_d$.  In MHD the magnetic field damping is usually
determined through the Reynolds number as $\xi_d/\xi_M=Re^{-3/4}$
\cite{dav}. The kinetic and magnetic  Reynolds numbers in the early
Universe can be extremely high, and thus one may expect that
$\xi_d\ll \xi_M$. In both helical and non-helical cases the
dissipative region of the energy density spectrum is given by a
Kolmogorov-type spectrum\footnote{By a Kolmogorov-type spectrum
we simply mean a $k^{-5/3}$ spectrum and ignore anisotropies
that are known to exist in non-helical MHD \cite{GS95}.
Such a spectrum can be derived from a phenomenological approach
too; see 5.3.2. of Ref.~\cite{b1}. Our numerical simulations
\cite{tkbk12,future} confirmed the Kolmogorov-type spectrum for a wide
range of magnetic Prandtl numbers} $E_M=C_K \varepsilon_M^{2/3} k^{-5/3}$,
where $C_K$ is a constant of order unity (1.6--1.7 for a wide
range of Reynolds numbers), $\varepsilon_M$ is the magnetic energy
dissipation rate per unit mass, given by $\varepsilon_M = 2\lambda
\int_{k_0}^{k_D} k^2 E_M(k)$ with $\lambda$ being the magnetic diffusivity.
At large scales above $\xi_M$ the magnetic field decay
is strongly dependent on the presence of magnetic helicity. The high
conductivity of plasma ensures magnetic helicity conservation that is
responsible for the transfer of spectral energy from small to
large scales via a so-called {\it inverse cascade}. In the non-helical
case the process is more complicated. As we will see below, magnetic
helicity conservations leads to a faster growth of the correlation
length and a slower decay of total magnetic energy.

\subsubsection{Non-helical magnetic fields}

As we already stated above, causality requires that $E_M(k) \propto k^4$
and this is a consequence of the divergence-free condition for the
magnetic field \cite{cd01}.
On the other hand, there is no zero-divergence requirement for the velocity
field, and this allows for the possibility to have a white noise spectrum
for the velocity field, i.e.\ $E_K (k) \propto k^2$; see \cite{hogan}.
We would like to note that our numerical simulations allowing the
longitudinal forcing, see \cite{kbtr10,kbcrt12}, show a white noise
spectrum for the velocity field as a final configuration.
Under these conditions, the power of magnetic field modes on the large
scales is much smaller than the power of plasma motions.
Thus, potentially the magnetic field might be  amplified via a transfer
of energy from plasma motions at large scales.
The time scale $\eta$ on which the field can be amplified at large
scales can be deduced from the induction equation
\begin{equation}
\frac{\partial B}{\partial t}\sim \frac{v_K(L) B}{L} \rightarrow \eta \sim L/v_L
\end{equation}
i.e., the characteristic field amplification time scale is approximately
the plasma eddy turnover time.
On this time scale an equipartition between kinetic and magnetic energies
could be reached over the distance range $L$.
In the final configuration the equipartition between magnetic and kinetic
energies is a consequence of the coupling between magnetic and
velocity fields.\footnote{In fact,
the two Reynolds numbers in the Universe are
high enough to ensure the validity of the Kolmogorov-type
phenomenological approach \cite{b2}, according to which there is
self-similarity between kinetic and magnetic energy densities
evolution (see Sec.~7.3.4 of Ref.~\cite{bII}), i.e.\
\begin{equation}
{\mathcal E}_M \sim {\mathcal E}_K \sim {\mathcal E},
\label{EM-EK}
\end{equation}
}
The growth of the magnetic field up to equipartition with the fluid
on large scales is somewhat similar to the phenomenon of an
``inverse cascade'' of the magnetic power spectrum.
Note however that the source of power in the inverse transfer of
non-helical magnetic fields is different from the power source in the
case of helical fields.
In the case of non-helical fields, the power in the large wavelength
modes increases due to the presence of a large power reservoir in the
form of the turbulent motions of the plasma in the same wavelength range.
By contrast, in the case of helical fields, the power on large scales
grows due to the transfer of power from the shorter wavelength modes.

Apart from the transfer of power from the plasma motions to the magnetic
field at large scales, another effect of evolution of plasma and magnetic
field perturbations is the turbulent decay of the power at short length
scales due to the phenomenon of the direct turbulent cascade.
At a given moment of time $\eta$, this phenomenon leads to suppression
of power in velocity and magnetic field modes on scales smaller than
the size of the largest processed eddy.
The size of the largest processed eddy is determined by the condition
\begin{equation}
\xi_K\sim\xi_M\sim v_K(\xi_K)\eta,
\end{equation}
where $v_L(\xi)$ is the characteristic velocity of the plasma motions on
the scale $\xi$.
From the definition of the kinetic energy power spectrum through the
velocity two-point correlation $P_K({\bf k})\sim |v_k|^2\sim |v_K(L\sim
1/k)|^2/k^2$ one finds that characteristic velocity $v_K$ on the
distance scale $L\sim k^{-1}$ is $v_K^2\sim k^2P_K$.
Since the power spectrum of plasma perturbations is $P_K\sim k^0$,
we have $v_K\sim k\sim \xi^{-1}$, so the above equation has the solution
\begin{equation}
\xi_K\sim\xi_M\sim \eta^{1/2}.
\end{equation}
The energy of plasma perturbations on the scale $\xi_K$ is $E_K=4\pi
k^2 P_K\sim k^2\sim \xi_K^{-2}\sim \eta^{-1}$.
Since the magnetic field on the scale $\xi_K\sim \xi_M$ is in equipartition
with the plasma, the energy density of the magnetic field evolves with time as
\begin{equation}
E_M\sim E_K\sim \eta^{-1},
\end{equation}
so that the strength of magnetic field evolves as
\begin{equation}
B^{\rm (eff)}\sim \sqrt{E_M}\sim \eta^{-1/2}.
\end{equation}

Below we demonstrate numerically (see Sec.~III) that the
evolution laws $\xi_M\sim \eta^{1/2}, E_M\sim \eta^{-1}$ are indeed
realized in the free turbulence decay regime. It should be
noted that the ``universality'' of the decay law $\xi_M \propto
\eta^{1/2}$ is not realized if we were to consider the magnetic field
evolution separately from the velocity field evolution \cite{b1};
see also Refs.~\cite{jedamzik,campanelli} for the magnetic field
decay laws in the cosmological context. Accounting for the Loitsianskii
invariant for turbulence leads to the decay laws being dependent on
the spectral shape; see Ref.~\cite{caprini-inv} and references therein.
It has also been claimed that the decay laws in the case of
non-helical magnetic fields strongly depend on the initial conditions
and can be different even when the helicity is extremely small \cite{b1}.

\subsubsection{Helical fields}

In the case of helical fields the evolution of $E_M$ and $\xi_M$
is determined directly by the condition of the conservation of magnetic
helicity, $ \int {\bm A}\cdot {\bm B}\, d^3x$.  High Reynolds numbers
allow us to follow the Kolmogorov-type phenomenological approach given
above; see Sec.~4.2.3 of Ref.~\cite{b1}.
Accounting for magnetic helicity conservation
${\mathcal E}_M \xi_M = {\rm const}$\footnote{Accounting for
magnetic helicity conservation and assuming that the
magnetic spectral energy is sharply peaked at $\xi_M$ a very rough
estimate implies that $B^{\rm eff} \propto \xi_M^{-1/2}$.} and
combining Eqs.~(\ref{EM-EK}) and $\xi_M \sim {\mathcal E}^{3/2}/ \varepsilon$,
which follows from the dimensionless analysis (based on the
Kolmogorov-type approach), we get \cite{b2}
\begin{equation}
-\frac{d{\mathcal E}_M}{d\eta} \sim {\mathcal E}_M^{-5/2},
\label{H}
\end{equation}
which leads to the decay laws ${\mathcal E}_M \propto \eta^{-2/3}$
and $\xi_M \propto \eta^{2/3}$. Below we
demonstrate numerically (see Sec.~\ref{IIIB}) the appearance of the $\xi_M\sim
t^{2/3}, E_M\sim t^{-2/3}$ laws in the evolution of helical magnetic
fields in the free decay regime.

\subsection{Simulations setup}

To model the evolution of the magnetic field and fluid perturbations
we solve the compressible equations with the pressure given by
$p=\rho c_s^2$, where $\rho$ is the gas density and $c_s=1/\sqrt{3}$
is the sound speed for an ultra-relativistic gas.
Following our earlier work \citep{kbtr10}, we solve the
governing equations for the logarithmic density $\ln\rho$,
the velocity $\bm{v}$, and the magnetic vector potential $\bm{A}$,
in the form
\begin{eqnarray}
\frac{\DD\ln\rho}{\DD\eta}&=&- \bm{\nabla}\cdot\bm{v},\\
\frac{\DD\bm{v}}{\DD\eta} &=&
{\bm{J}\times\bm{B} \over \rho}-c_s^2\bm{\nabla}\ln\rho +\bm{f}_{\rm visc}
\\
\frac{\partial\bm{A}}{\partial \eta}&=&
\bm{v}\times\bm{B}
+{\bm f}_M+\lambda \nabla^2{\bm A},
\label{mhd3}
\end{eqnarray}
where $\DD/\DD\eta=\partial/\partial\eta+\bm{v}\cdot\bm{\nabla}$
is the advective derivative, $\bm{f}_{\rm visc}=\nu\left(\nabla^2{\bm v}
+{\textstyle\frac{1}{3}}\bm{\nabla}\bm{\nabla}\cdot\bm{v}+\bm{G}\right)$
is the viscous force in the compressible case with constant kinematic
viscosity $\nu$ and $G_i=2{\sf S}_{ij}\nabla_j\ln\rho$ as well as
${\sf S}_{ij}=\frac{1}{2}(v_{i,j}+v_{j,i})-\frac{1}{3}\delta_{ij}v_{k,k}$
being the trace-free rate of strain tensor.
Furthermore, $\bm{J}=\bm{\nabla}\times\bm{B}/4\pi$ is the current density.

We use the {\sc Pencil Code} \cite{pencil} with a resolution of
$512^3$ meshpoints.
An important difference of our simulations
setup from those of previous studies is in the treatment of the
backreaction of fluid perturbations onto the magnetic field. Such treatment is
important, especially on large length scales, because the spectrum
of velocity perturbations follows a white noise ($E_K \propto k^2$)
spectrum to large scales \cite{hogan}. Large-scale fluid
perturbations affect the magnetic field evolution at the largest scales
and could lead to a transfer of power to the large-scale modes of the
magnetic field, an effect similar to the inverse cascade developing
even in the case of non-helical fields; see also
Sec.~\ref{sec:simulations}.

Prior to the simulation of the magnetic field decay we inject
magnetic energy into the computational domain at scales
corresponding to the phase transition eddy size see
Ref.~\cite{kbtr10} for more details. We approximate the magnetic
field injection by a delta function in wave
number space, allowing it to interact through MHD
processes with a rest plasma. After several turnover times the
initial sharp peak of the magnetic field starts to disappear and the
magnetic field begins to spread over a wide range of the wavenumbers;
see Fig.~4 of Ref.~\cite{kbtr10}. In a few turnover times the
spectrum becomes established with a cut-off at small length scales
and a well defined Kolmogorov-like integral scale {$k_M$}, and a $k^4$
spectral shape at large length scale. In contrast to the studies of
Refs.~\cite{jedamzik,campanelli} we recover the spectral shape of
the magnetic field at large scale.{\footnote{Note that in
Ref.~\cite{kbtr10} the spectral index between 3 and 4 due to a
different choice of the system parameters and run time.}} This is in
perfect agreement with previous analytical results of
Ref.~\cite{cd01} based on causality and divergence free field
arguments.

We use simple power-law models for the decay of turbulence and scale
the magnetic correlation length and magnetic field with
temperature as follows:
\begin{equation}
{\xi_M \over \lambda_0} = \left( {T \over T_\star} \right)^{-n_\xi}
~, \label{evolution-xi-nonhelical}
\end{equation}
\begin{equation}
{B^{\rm (eff)} \over B_\star} = \left( {T \over T_\star} \right)^{-n_E} ~,
\label{evolution-b-nonhelical}
\end{equation}
where $B_\star$ and $T_\star$ are the effective values of the
magnetic field at the moment of generation and the temperature of
the phase transition, respectively. Hence, the values of the parameters
$n_\xi$ and $n_E$ describe the turbulent decay laws that differ from
each other in the non-helical and helical cases.

Qualitative arguments presented above show that the expected
values for $(n_\xi, n_E)$ for non-helical and helical fields are
$(1/2, -1/2)$ and $(2/3,-1/3)$, respectively. Below we show that this is
indeed the case.

\section{Simulations results}
\label{sec:simulations}

In Sec.~II we have presented a phenomenological description of the scaling laws for
the magnetic correlation length and the magnetic energy in the
free turbulence decay regime. Below we address the same scaling laws
based on our simulations. We also briefly review the results of previous works.

\subsection{Non-helical Magnetic Field Evolution}

\begin{figure}[t]
\includegraphics[width=\columnwidth]{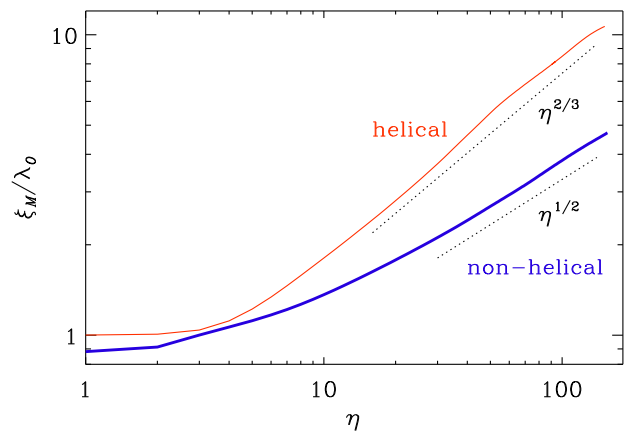}
\caption{(color online).
$\xi_M(\xi)$ for helical (thin, red) and non-helical (thick, blue) cases.
}\label{pcomp_kf}\end{figure}

\begin{figure}[t]
\includegraphics[width=\columnwidth]{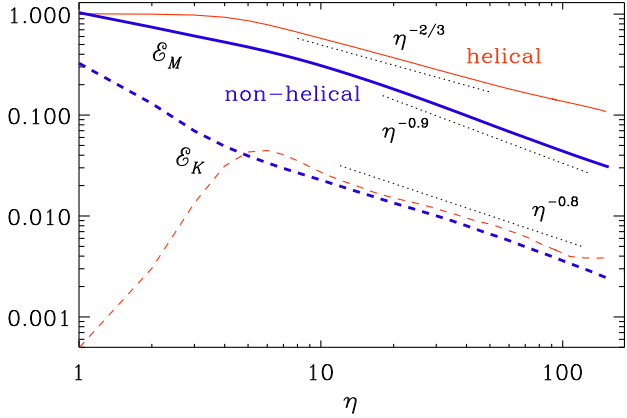}
\caption{(color online).
${\cal E}_M(\xi)$ (solid) and ${\cal E}_K(\xi)$ (dashed)
for the helical (thin, red) and non-helical (thick, blue) cases.
}\label{pcomp_EM_EK}\end{figure}

\begin{figure}[t]
\includegraphics[width=\columnwidth]{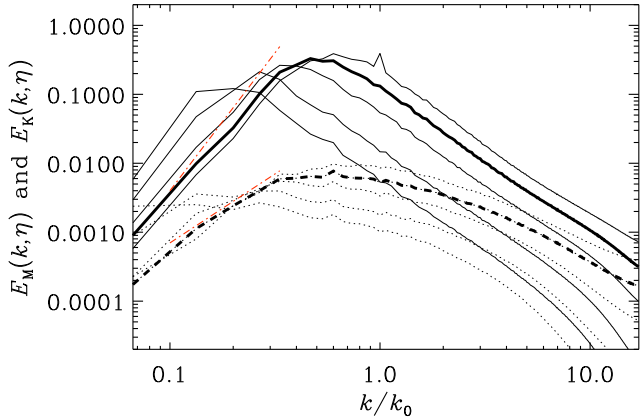}
\caption{(color online).
Evolution of $E_M(k,\xi)$ (solid) and $E_K(k,\xi)$ (dashed)
versus $k$ for $\eta=5$, 10, 20, 50, and 100 for the non-helical run.
Thick lines are for $\eta=10$.
The red dash-dotted lines give the $k^2$ and $k^4$ scalings for comparison.
All spectra are normalized by $\int E_M(k,0)\,dk/k_0$.
}\label{pkt512_short512pm1b3_noforce}\end{figure}

The scaling laws for the non-helical magnetic field evolution have
been studied through different simulations by different groups; see
Refs.~\cite{a1,s1,b2,jedamzik,maclow} and references therein. As is
stated in Ref.~\cite{b1}, the magnetic decay laws for the
non-helical case strongly depend on initial conditions, and result
in exponents $n$ in the decay law ${\cal E}_M (\eta) \propto
\eta^{-n}$ that vary in the range $1.3>n>0.65$. Note that the
numerical and phenomenological studies performed in
Refs.~\cite{jedamzik,campanelli} lead to $\xi_M(\eta ) \propto
\eta^{0.4}$ and ${\mathcal E}_M (\eta ) \propto \eta^{-1.2}$ for a
white noise spectrum, and this is in good agreement with the grid
turbulence description of hydrodynamic turbulence \cite{dav}.
On the other hand, the 3D MHD
simulations of Refs.~\cite{a1,kbtr10} and the phenomenological study
of Ref.~\cite{campanelli2} show a slightly faster growth of the
correlation length $\xi_M (\eta) \propto \eta^{1/2}$ with a magnetic
energy decaying as ${\mathcal E}_M \propto t^{-1}$. The
difference between two different scaling laws is probably due to different
initial conditions. In particular, the initial velocity field has
traditionally been taken to be zero. By contrast, here we have taken
as initial condition the result of a self-consistent magnetically
driven turbulence simulation. The numerical simulations of
Ref.~\cite{future} show that the growth of the correlation length is
almost independent of the magnetic Prandtl number with {$\xi_M
\propto \eta^{1/2}$} (Fig.~\ref{pcomp_kf}), while the exponent of
the total magnetic energy density decay is compatible with $-1$
(here closest to $-0.9$; Fig.~\ref{pcomp_EM_EK}). Accounting for $T
\propto a^{-1}$ and $\propto \eta^{-1}$ during the radiation
dominated epoch, and the magnetic field strength $B^{\rm
(eff)}=\sqrt{8\pi \mathcal E}_M$ we get the scaling indices for the
decay of non-helical turbulence: $n_\xi = 1/2$ and $n_E = -1/2$.
As expected, for $k\ll k_0$ early times, we find
$E_M(k,\xi)\propto k^4$ and $E_K(k,\xi)\propto k^2$;
see \Fig{pkt512_short512pm1b3_noforce}.
Furthermore, even in this non-helical case the spectral energies
increase with time for $k\ll k_0$, while for $k\gg k_0$ they decrease.

\subsection{Helical Magnetic Field Evolution}
\label{IIIB}

As we have noted above, the presence of magnetic helicity results in the
development of an inverse cascade during which the correlation
length is increasing while the total magnetic energy decreases.
Similar to the non-helical case there are basically two different
approaches: (i) Refs.~\cite{b2,jedamzik,campanelli} assume exact
conservation of magnetic helicity and the magnetic field is the dominant
contribution to the total energy density, i.e., ${\mathcal
E}_K/{\mathcal E}_M \ll 1$ (where ${\mathcal E}_K$ is the total
kinetic energy density of turbulence); (ii) other approaches are
given in Refs.~\cite{a1,campanelli2}. In particular, Ref.~\cite{a1}
refers to a more general case with magnetic helicity evolving as ${\mathcal
H}_M (\eta) \propto \eta^{-2s}$. Also, the ratio between kinetic and
magnetic energy densities has in some studies assumed to be around
1; Ref.~\cite{campanelli2} assumes that the magnetic field evolves
toward a force-free regime with constant magnetic helicity and with
a constant ratio between magnetic and kinetic energy densities. All
models \cite{s1,b2,campanelli2,jedamzik,campanelli} show that the
scaling laws are independent of the initial magnetic field spectrum.
In fact, there are two main behaviors described:
(i) Refs.~\cite{b2,jedamzik,campanelli} claim ${\mathcal E}_M (\eta) \propto
\eta^{-2/3}$ and $\xi_M (\eta) \propto \eta^{2/3}$;
(ii) Refs.~\cite{a1,campanelli2} claim  $\xi_M (\eta) \propto \eta^{1/2}$;
with ${\mathcal E}_M (\eta) \propto \eta^{-1/2}$. In both scaling laws
$\xi_M {\mathcal E}_M \sim {\rm const}$. The main difference between
these two scaling laws consists in choosing the turbulence model.
Refs.~\cite{a1,campanelli2} assume a force-free development of the MHD
turbulence decay, while Ref.~\cite{campanelli}, see their Eq.~(4),
assumes a linear dependence between vorticity and Lorentz force.
Our new numerical results support the former scenario (i).
Fig.~\ref{pkt512_short512pm1b3_noforce} shows the evolution of
kinetic and magnetic spectral energies. As we can see, the $k^2$
and $k^4$ laws are established at large scales for kinetic and
magnetic spectral energies, respectively. We can also see the
slight increase of power at large scales, even in the case of non-helical fields.

\begin{figure}[t]
\includegraphics[width=\columnwidth]{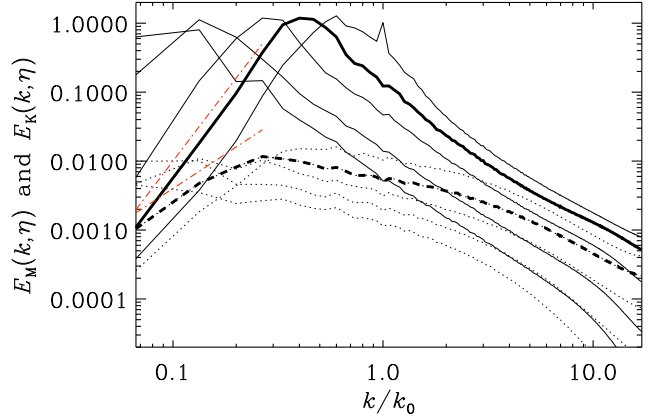}
\caption{(color online).
Evolution of $E_M(k,\xi)$ (solid) and $E_K(k,\xi)$ (dashed)
versus $k$ for $\eta=5$, 10, 20, 50, and 100 for the helical run.
Thick lines are for $\eta=10$.
The red dash-dotted lines give the $k^2$ and $k^4$ scalings for comparison.
All spectra are normalized by $\int E_M(k,0)\,dk/k_0$.
}\label{pkt512_short512pm1a3_noforce}\end{figure}

We have performed a study of the large-scale decay of a maximally
helical magnetic field under conditions similar to those in the
non-helical case, and for different magnetic Prandtl numbers as well
as different values of magnetic resistivity. We have recovered the
$E_M(k) \propto k^4$ spectral shape and the scaling laws as $\xi_M
\propto \eta^{2/3}$ and ${\mathcal E}_M \propto \eta^{-2/3}$, so
that $n_\xi=2/3, n_E=-1/3$ \cite{BM99}; see also \cite{b1} for more
general discussion; see Figs.~\ref{pcomp_kf} and
\ref{pcomp_EM_EK}. Again these scaling laws are valid when the
correlation length is greater than the damping scale, so that
dissipation does not play an important role.
Similarly to the non-helical run, we find for $k\ll k_0$ and early times
that $E_M(k,\xi)\propto k^4$ and $E_K(k,\xi)\propto k^2$;
see \Fig{pkt512_short512pm1a3_noforce}.
In this case there is a strong inverse cascade with a strong
increase of spectral energies with time for $k\ll k_0$.
We can also see the constant magnetic power while the peak
is moving toward large scales (inverse cascade).
This corresponds to the constant helicity case.

\begin{figure}[t]
\begin{center}
\includegraphics[width=0.98\columnwidth]{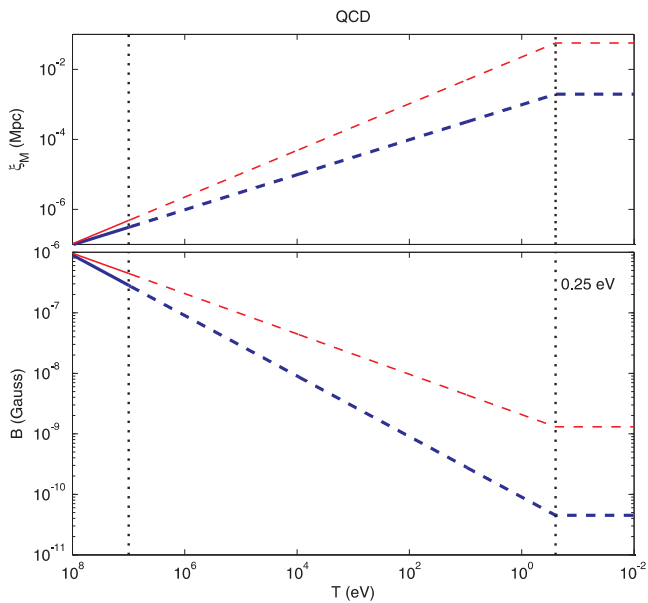}
\end{center}
\caption{(color online).
The correlation length $\xi_M$ (top panel) and the maximal
allowed $B_{\rm max}$ (bottom panel) for a primordial helical (thin red)
and non-helical (thick blue) magnetic fields generated during QCDPT.
Constraints on the magnetic field at $T=0.25$eV are set to
$B_{\rm max}=4.5\times 10^{-11}$Gauss (${\xi_M=2\times 10^{-3}{\rm Mpc}}$)
and $B_{\rm max}=1.3\times 10^{-9}$Gauss (${\xi_M=5.6\times 10^{-2}{\rm Mpc}}$)
for non-helical and helical cases, respectively. Dashed lines show areas where
damping processes may reduce ideal estimates.}
\label{QCD}
\end{figure}

\begin{figure}[t]
\begin{center}
\includegraphics[width=0.98\columnwidth]{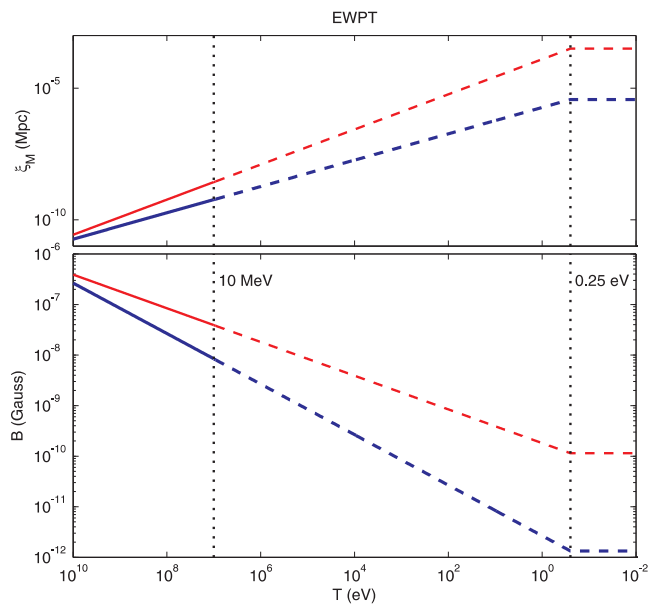}
\end{center}
\caption{(color online).
The correlation length $\xi_M$ (top panel) and the maximal
allowed $B_{\rm max}$ (bottom panel) for a primordial helical (thin red)
and non-helical (thick blue) magnetic fields generated during EWPT.
Constraints on the magnetic field at $T=0.25$eV are set to
$B_{\rm max}=1.3\times 10^{-12}$Gauss ($\xi_M=3.5\times 10^{-6}{\rm Mpc}$)
and $B_{\rm max}=1.1\times 10^{-10}$Gauss ($\xi_M=3.1\times 10^{-4}{\rm Mpc}$)
for non-helical and helical cases, respectively. Dashed lines show areas where
damping processes may reduce ideal estimates.}
\label{EWPT}
\end{figure}

\section{Implication for cosmological evolution of magnetic field}

Decay of cosmological MHD turbulence occurs together with the
cooling of the Universe and the increase of magnetic
correlation length. Therewith, the correlation length increases only up
to the point when the Universe reaches the temperature $T=1{\rm\ eV}$
\cite{jedamzik2}. The initial values of correlation length and magnetic
field strength, $\xi_0$ and $B_0$, at the temperature of magnetogenesis $T_*$,
together with the two scaling indices $n_\xi$ and $n_E$, fully determine the
large-scale magnetic field decay, and as a result the final
configuration of the magnetic field.

Phenomenological arguments as well as numerical simulations show that
$n_\xi=1/2$ and $n_E=-1/2$ in non-helical case, while $n_\xi=2/3$
and $n_E=-1/3$ in the helical case during the turbulent regime.
The speed of growth of $\xi_M$ is constant, independently of the relation
between $\xi_M$ and $\xi_d$. This implies that, in the case in which the
initial correlation length of magnetic field is comparable to the
size of cosmological horizon at the epoch of magnetogenesis, the
final correlation length reaches $2 \times 10^{-4}$ Mpc and $6
\times 10^{-3}$ Mpc for non-helical magnetic fields generated at
EWPT and QCDPT, respectively. In the helical case, the correlation length
reaches 10\,kpc for fields generated at the QCDPT and $0.1$
Mpc for the EWPT.

\begin{figure}[th]
\includegraphics[width=\linewidth]{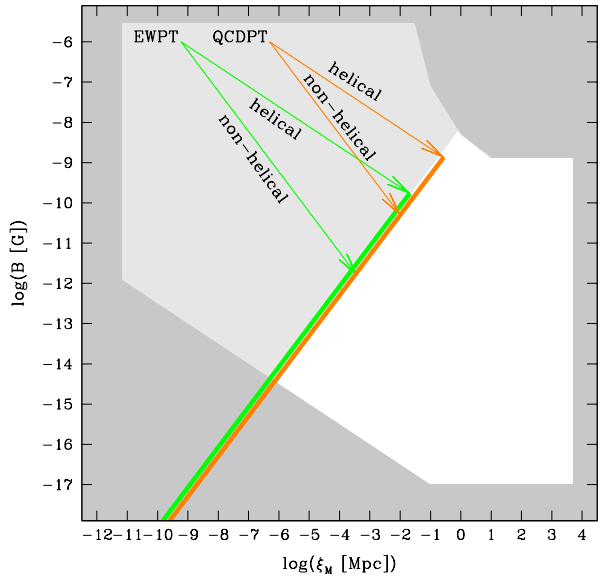}
\caption{(color online).
Cosmological evolution of $B^{\rm (eff)}$ and $\xi_M$ for magnetic fields generated at the EWPT (green) and QCDPT (orange).  Arrows show the evolution of the strength and integral scale of helical and
non-helical fields during the radiation dominated era up to their final values. Thick solid line(s) show possible present day strength and integral scale of the phase-transition generated magnetic
fields.}
\label{fig:nonhelical}
\end{figure}

The evolutionary paths of magnetic field strength and correlation
length in time and in the $B,\xi_M$ parameter space are shown in
Figs.~\ref{QCD}--\ref{fig:nonhelical}.
In principle, the free turbulence decay periods in the early Universe
are intermittent, with periods of viscously damped evolution
\citep{jedamzik}.
So, starting from 10\,MeV, Figs.~\ref{QCD} and \ref{EWPT} show only maximal
values of helical and non-helical magnetic fields and their correlation length
from QCD and EW phase transitions without accounting for any possible damping
processes that can affect the given scaling laws.
These maximal values are derived using several assumptions: the
magnetic correlation length during phase transition matches the
bubble size, and magnetic fields are excited with maximal amplitudes
allowed by the BBN limit.
In fact, before neutrino decoupling the viscous damping force
${\bf f_{\rm visc}}$ in Eq.~(14) grows as
$\nu_\sim l_{\rm mfp,\ \nu}\sim \eta^4$, where $l_{\rm mfp,\ \nu}$
is the neutrino mean free path.
This growth is much faster than the $\xi_M\sim \eta^{1/2}$ growth
of the integral scale of the magnetic field.
This means that, even if $\xi_M\gg l_{\rm mfp,\ \nu}$ at the moment of
magnetogenesis,  $l_{\rm mfp}$ catches up with $\xi_M$ at a later time
$\eta_{\rm visc}$.
Starting from this time and up to the moment of neutrino decoupling,
the magnetic field stops to decay, $B\sim{\rm const}$, because the
fluid motions are damped by viscosity, $v\sim 0$, so that there is
no coupling of magnetic field to the fluid in this regime.
However, turbulence re-starts after the neutrino decoupling, so
that the system returns to the same evolutionary track shown in
Fig.~\ref{fig:nonhelical} just after neutrino decoupling.
At lower temperatures, when the viscosity is provided by photon streaming,
the viscous damping scale grows as $\nu\sim l_{\rm mfp,\ \gamma}\sim\eta^2$,
where $l_{\rm mfp,\ \gamma}$ is the photon mean free path.
In this time interval the growth of $\nu$ is again faster than the growth
of $\xi_M$, so that the episode of viscously damped evolution repeats
when $l_{\rm mfp,\ \gamma}$ reaches $\xi_{M}$.
This could again delay the advance of $(B, \xi_B)$ along the evolutionary
track shown in Fig.~\ref{fig:nonhelical}.
The end point of the evolutionary track at the end of the
radiation-dominated era is well defined by the condition that the
correlation length of magnetic field should not be shorter than the Silk
damping scale times the Alfv\'en velocity \citep{va}.
The loci of the possible end points of the evolution are shown by the
inclined thick solid (green/orange) lines in Fig.~\ref{fig:nonhelical}.

\section{Conclusion}

Our study shows that magnetic fields generated during phase
transitions are comparable with the observational lower bound even
if we account for large-scale decay as well as additional Alfv\'en
wave-induced damping. The extremely low values of the
smoothed magnetic field \cite{caprini01} do not imply that the
effective magnetic field in the range $1\,$pc--$1\,$kpc are small enough to
result in observational changes in blazar emission spectra.
The advantage of using the effective magnetic field lies in its
independence of the spectral shape.
In summary, if the magnetic field has been generated
during a phase transition, its correlation length is strongly
limited. If future observations were to detect a weak magnetic
field $\leq  10^{-14}$--$10^{-15}$\,G with a typical correlation length
of the order of a few pc, this could serve as an indication of
magnetogenesis during EWPT, while a somewhat stronger field with a
correlation length of the order of kpc might indicate the presence of
QCDPT magnetogenesis.
\\

\acknowledgments  We acknowledge fruitful discussions with R.\ Durrer.
We appreciate helpful comments from L.\ Campanelli,
K.\ Jedamzik, A.\ Kosowsky, and B.\ Ratra. We acknowledge partial
support of computing resources provided by the Swedish
National Allocations Committee at the Center for Parallel Computers
at the Royal Institute of Technology in Stockholm and the Carnegie
Mellon University supercomputer center. We acknowledge partial
support from Swiss National Science Foundation SCOPES grant no.\
128040, NSF grant AST1109180 and NASA Astrophysics Theory Program
grant NNXlOAC85G. This work was supported in part by the European
Research Council under the AstroDyn Research Project 227952 and the
Swedish Research Council grant 621-2007-4064. T.K.\ acknowledges the
ICTP associate membership program. A.B.\ and A.T.\ acknowledge the
McWilliams Center for Cosmology for hospitality.

{}
\end{document}